\documentclass[journal,a4paper]{IEEEtran}
\usepackage{graphicx} 
\usepackage{amsmath}
\usepackage{float}
\usepackage{subfig}
\usepackage{caption}
\usepackage{url}
\usepackage{hyperref}

\newcommand{\figwidth}{0.98\columnwidth}

\usepackage[boxed,ruled,vlined,linesnumbered,commentsnumbered, noresetcount]{algorithm2e}
\usepackage{algorithmic}

\usepackage{diagbox}
\usepackage{tikz}
\usetikzlibrary{shapes.geometric}
\usetikzlibrary{positioning}

\usepackage{flushend}

\usepackage{tabularx}
\usepackage{ragged2e} 
\newcolumntype{L}{>{\RaggedRight}X} 
\usepackage{lipsum} 
\usepackage{amsmath}
\usepackage{amssymb}
\usepackage{comment}

\makeatletter
\newcommand\notsotiny{\@setfontsize\notsotiny\@vipt\@viipt}
\makeatother

\begin{document}


\title{Industry 4.0 and Beyond: The Role of 5G, WiFi 7, and TSN in Enabling Smart Manufacturing}


\author{Jobish John, Md. Noor-A-Rahim,  Aswathi Vijayan,  H.~Vincent~Poor~\IEEEmembership{Life Fellow IEEE},   and Dirk Pesch~\IEEEmembership{Senior Member IEEE}


\thanks{Jobish John is with the CWTe, Department of Electrical Engineering, Eindhoven University of Technology, The Netherlands. The work reported in the paper was carried out while Dr. John was a research fellow at UCC. 
(E-mail: {\tt j.john@tue.nl}).  

Md. Noor-A-Rahim and   Dirk Pesch  are with the  School of Computer Science \& IT, University College Cork,  Ireland.  (E-mail: {\tt \{j.john, m.rahim, d.pesch\}@cs.ucc.ie}). 

Aswathi Vijayan is with the  School of Electronics Engineering, VIT Vellore. (E-mail: {\tt aswathi.v@vit.ac.in}).

H. V. Poor is with the  Department of Electrical and Computer Engineering, Princeton University, USA (E-mail: {\tt poor@princeton.edu}).
}}


\maketitle

\begin{abstract}
This paper explores the role that 5G, WiFi-7, and Time-Sensitive Networking (TSN) can play in driving smart manufacturing as a fundamental part of the Industry 4.0 vision. The paper provides an in-depth analysis of each technology's application in industrial communications, with a focus on TSN and its key elements that enable reliable and secure communication in industrial networks. In addition, the paper includes a comparative study of these technologies, analyzing them based on a number of industrial use-cases, supported secondary applications, industry adoption, and current market trends. The paper concludes by highlighting the challenges and future directions for the adoption of these technologies in industrial networks and emphasizes their importance in realizing the Industry 4.0 vision within the context of smart manufacturing.
 
\end{abstract}

\begin{IEEEkeywords}
Industrial IoT, Industry 4.0, Industry 5.0, cyber-physical manufacturing systems, Millimeter-wave,  Smart Manufacturing, Smart Factory, Wireless Factory, 5G, TSN, 6G, WiFi-7, WiFi-8, IEEE 802.11be,  Wireless TSN.

\end{IEEEkeywords}



\section{Introduction}
\begin{figure*}[htbp]
\centering
\includegraphics[width=6.5 in, height=2.75 in] {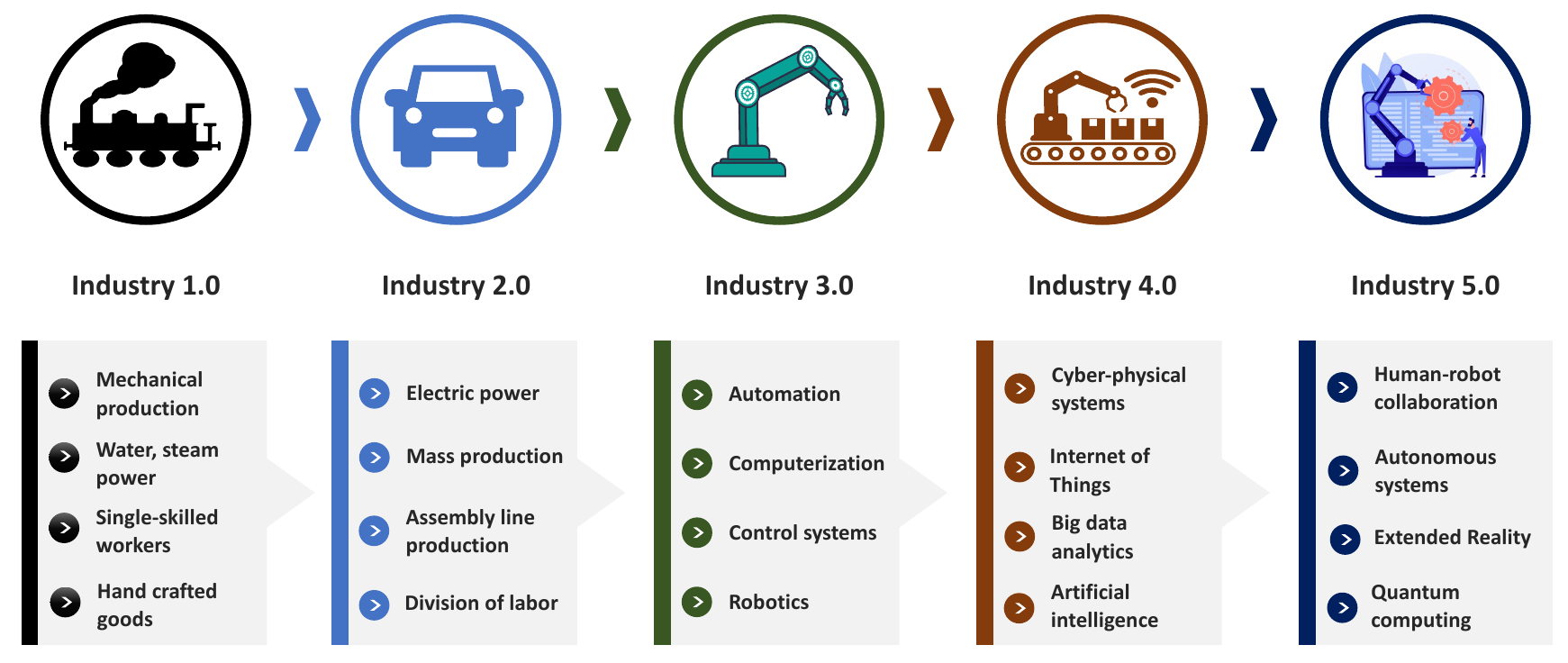}
\caption{Industrial evolution in the context of smart manufacturing.}
\label{fig:evo}
\end{figure*}

Smart manufacturing is a revolutionary concept  that aims to transform the manufacturing sector with the utilization of modern technologies \cite{Fei_2019}. The objective of smart manufacturing is to create a more productive, flexible, and eco-friendly manufacturing system that can adapt to the changing market needs and, consequently, boost the overall performance of manufacturing processes. The successful implementation of smart manufacturing depends on the effective use of industrial communications (also known as industrial networking). It allows different machinery, systems and devices to exchange data in a timely and uninterrupted manner, which is critical for the well-coordinated execution of manufacturing processes,  production planning, quality assurance and maintenance tasks. This combination of physical and digital systems is a central part of the fourth industrial revolution (otherwise known as Industry 4.0), which is transforming conventional industrial automation and control systems into sophisticated cyber-physical manufacturing systems. Beyond Industry 4.0, we expect to see an environment where autonomous machines and humans collaborate to achieve common goals. This further evolution towards what might be called Industry 5.0, will leverage new technologies such as augmented and extended reality (A/XR), autonomic systems, and 6G networks \cite{Our_IRS} to create a more autonomous, self-adjusting, and adjustable manufacturing system that can foresee and respond to fluctuating market needs. The industrial evolution in the context of smart manufacturing is summarized and illustrated in Fig~\ref{fig:evo}.

 The recent developments in communication technologies, such as 5G, WiFi-7 and Time-Sensitive Networking (TSN), are expected to be one of the key enablers of smart manufacturing in the Industry 4.0 and beyond era.  These technologies have the potential to  completely change the way industrial devices communicate with the digital world, allowing a vast  number of industrial machines to be connected to the Internet, interacting with an array of IT applications that are used to manage industrial organisations, thus enabling the  development of faster, more efficient, and more responsive manufacturing systems.

The introduction of 5G networks has opened up new possibilities for wireless communication, especially in the field of smart manufacturing. 5G has specified an advanced cellular network that offers significant improvements over their predecessors in terms of speed, capacity, and low latency to support not only traditional mobile broadband services but a wide range of machine-to-machine communication services \cite{Chettri_M2M}.  The high-speed and low-latency capabilities of 5G, combined with features such as the New Radio interface,   network slicing, and software-defined networking (SDN), is expected to provide the foundation for a highly connected and efficient industrial ecosystem. 5G's advanced error correction codes, mmWave, massive MIMO, and improved signal processing further enhance reliability and performance, making it possible to connect a large number of devices and sensors in real time. These capabilities of 5G make it an ideal solution for smart manufacturing and Industry 4.0, allowing for the seamless exchange of data between machinery, systems, and devices.

WiFi 7, the latest development in wireless local area network technology, boasts an array of technical features such as the use of higher order modulation (4096 QAM), multi-link operation, wider bandwidth, multi-AP Operation, and WiFi Sensing, providing enhanced connectivity and increased efficiency needed in Industry 4.0 and beyond \cite{8847238}. The high modulation order enables fast data transmission, while multi-link operation allows a device to connect to the network using multiple links, thereby enhancing reliability, performance, and capacity. The extended bandwidth of WiFi 7 permits more data to be sent, thus reducing latency and enhancing effectiveness. Multi-AP Operation, by operating numerous access points as one entity, furnishes uninterrupted coverage even in large facilities, while WiFi Sensing offers integrated location-based services and upgraded asset tracking.

Time-Sensitive Networking, a set of standards under development by the IEEE 802.1 working group,  facilitates the development of smart manufacturing. This technology includes some unique features such as time synchronization, bounded end-to-end latency, dependability, and resource/network management which guarantee real-time communication and control between devices \cite{TSN_survey_Seol}. Fine-grained time synchronization across all devices on the network allows these devices to collaborate, while bounded end-to-end latency guarantees an expected response time. Additionally, TSN's high reliability makes sure that communication and control systems remain in working order even when network failures occur. Furthermore, resource and network management offer an effective distribution of network resources, giving priority to critical data transmission. Wireless TSN takes these advantages to a higher level, enabling real-time communication and control for mobile devices and machines, in particular in places where wired solutions are not practical.


The objective of this  paper is to examine the role of 5G, WiFi-7, and TSN in driving smart manufacturing in the Industry 4.0 and beyond era. We provide a comprehensive overview of the key enabling features of these technologies and their applicability in the context of future smart manufacturing.  The paper will also provide a comparison of the technologies and their impact on smart manufacturing use-cases and will discuss the trade-offs and potential for synergies between these technologies. To stimulate future research, we discuss a number of challenges and promising future research directions associated with the future smart manufacturing landscape.

The rest of this paper is structured as follows: In Section~\ref{sec_private5g}, we provide an overview of the key technologies in 5G private networks that are specifically designed for industrial communications. Section~\ref{sec_wifi7} discusses the key features of WiFi-7 and its relevance in industrial settings. Section~\ref{sec_tsn} covers the characteristics of TSN and provides details on TSN over wireless networks such as WiFi and 5G. A comparative analysis of these communication technologies is presented in Section~\ref{sec:Chall}. In Section VI, we highlight the challenges and future directions of industrial communication technologies. Finally, we conclude this paper in Section~\ref{sec:Con}.

\section{5G for Industrial Communications}
\label{sec_private5g}






 With the evolution from 4G to 5G, more varied usage scenarios and applications for cellular networks will be supported beyond voice, messaging and mobile broadband. 5G will support three service categories, that is enhanced mobile broadband (eMBB), ultra-reliable low-latency communications, massive machine type communications as well as fixed wireless access, which the International Telecommunication Union's Radio communication Sector (ITU-R) has identified for the 5G era \cite{Chettri_M2M}. By integrating 5G networks into their information and communication systems, enterprises can enhance their systems' and operations' performance by taking advantage of the enhanced capabilities of 5G networks. Some of them are listed below:

\begin{table*}
\centering
  \caption{Manufacturing use cases enabled by 5G.}
  \label{TABLE}
  \includegraphics[width=16 cm]{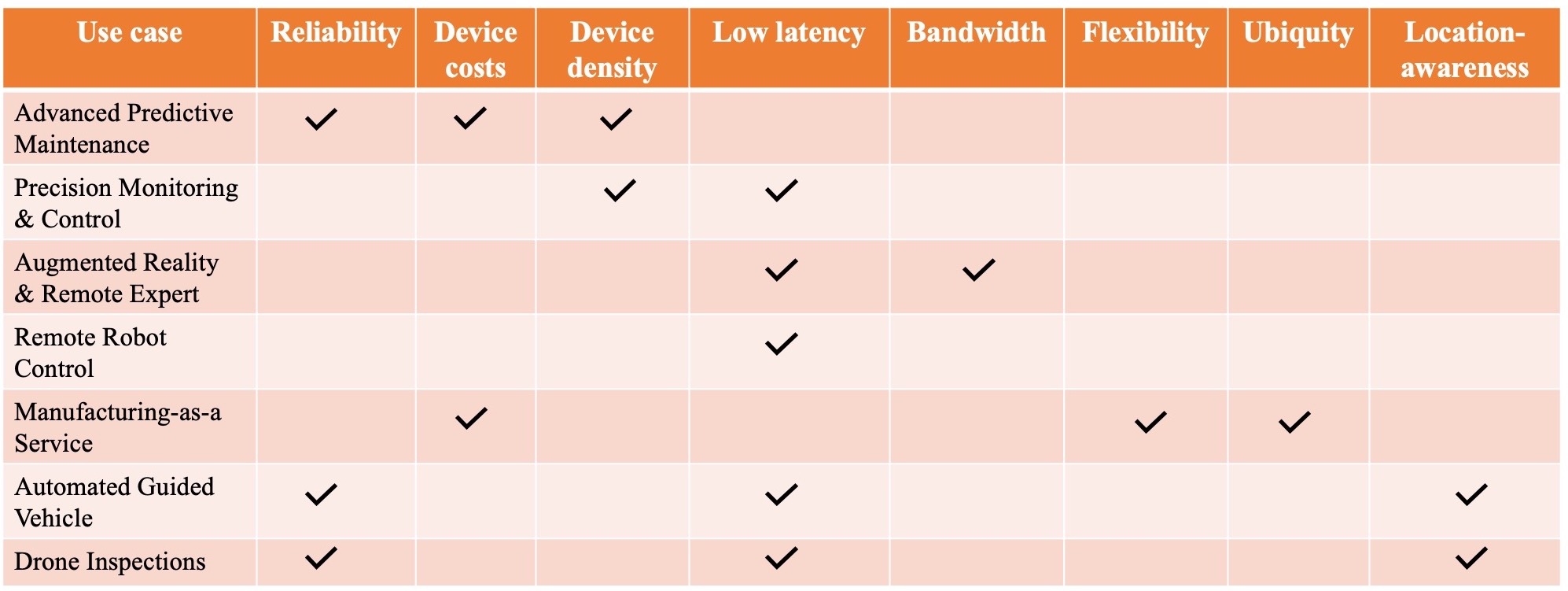}
\end{table*}
 
 \begin{itemize}

 \item \textbf{Flexible Numerology} for radio resource allocation. 3GPP outlines two main frequency ranges for 5G NR use, which are called frequency range 1 (FR1) and frequency range 2 (FR2). FR1 is also known as the sub6 GHz band, while FR2 is referred to as the millimetre wave (mmWave) band. The maximum channel bandwidth and the space between OFDM subcarriers can vary depending on the specific frequency range being used.  The concept of flexible numerology allows for variations in both the value of sub-carrier spacing and the duration of OFDM symbols, which impact the available data rate and transmission latency. \\
 These sub-carrier spacings  are obtained from scaling up the LTE based sub-carrier spacing $\delta f = 15 \mbox{kHz}$ by $2^\mu$. leading to a range from 15~kHz up to 240~kHz with a proportional change in cyclic prefix duration. This can offer the benefits of tailoring radio access parameters to suit the unique demands of industrial applications, like achieving low latency for real-time control or high throughput for data-intensive tasks. Furthermore, the adaptability of numerology allows for different applications with varying requirements to operate together within the same frequency band, resulting in optimal spectrum usage. This can lead to enhanced efficiency and productivity in manufacturing processes, as well as improved automation and quality control beyond what other wireless technologies can offer.

 \item \textbf{mmWave} communication: The allocated mmWave radio spectrum provides much more bandwidth than is available at sub-6GHz, permitting to accommodate a wide range of novel applications for Industry 4.0 and beyond. Example applications include advanced smart industrial functions like vision-guided robots, ultra-high definition video and imaging for remote visual monitoring and inspection, smart safety instrumented systems, intelligent logistics, and high precision image-guided automated assembly, among others. The availability of Ultra-Reliable and Low-Latency Communications (URLLC) in factory automation scenarios enables smart machines and robots to work alongside humans or cooperate towards a common goal, which is a key aspect of the future Industry 5.0 vision. Furthermore, utilizing mmWaves, enables not high throughout communication but also sensing, which can enable seamless and adaptive behaviour in equipment and machines, allowing them to detect adjacent individuals or objects and react appropriately by adjusting their movements or slowing their operating rate.
 
 \item \textbf{Beamforming}: Antenna beamforming employs an array of multiple antenna elements to generate a directed beam. This has the significant advantage of reducing interference in  sub-6-GHz bands, resulting in higher throughput due to directional transmission. At mmWave frequencies, beamforming is essential for reliable communication as it enhances channel gain. Beamforming, for example, facilitates concurrent communication among collaborative robots in a smart industrial environment.
 
 \item \textbf{Massive MIMO}: Massive MIMO utilizes a large number of antennas to exploit  spatial degrees of freedom, enabling it to support communication with multiple devices simultaneously without requiring additional time or frequency resources \cite{Zhang_2020}. As industrial environments tend to be full of metallic surfaces from equipment, many such environments are challenging for wireless communication. Massive MIMO's channel hardening effect improves its immunity to fast fading and allows for more deterministic communications, important for many industrial applications with strict quality of service requirements. 

     
 \item \textbf{Network Slicing}: Next-generation factories will need to handle diverse traffic flows that may have conflicting needs for performance, reliability, and security. A single large system cannot meet the demands of these new industrial situations. 
 Slicing permits the delivery of a variety of specific services with potentially incompatible requirements on a single physical 5G substrate  \cite{Ordonez_2018}.

 \item \textbf{5G LAN-type Service}: Most current automation systems in industry are based on a range of proprietary wired local area network (LAN) technologies. This allows devices to communicate directly with each other across the LAN, discover their services, utilise multi-cast communication and other LAN features. This is in contrast to 5G communication modes, which are more peer-to-peer oriented and rely on switching and routing in the 5G core network. The 5G LAN-type service \cite{rel16-lan} is designed to replicate the LAN features and simplify communication between 5G based devices, in particular in industrial automation environments \cite{5glan}. 

\end{itemize}


The manufacturing industry is rapidly adopting 5G wireless communication technology to automate industrial processes. 
The 5G network must be connected with the industrial facilities and equipment when it is deployed in an industrial manufacturing facility. Typically, a dedicated, private 5G network that is tailored to the requirements and services of the plant is needed. It can be difficult to get good radio reception in a manufacturing setting. Particularly in large factories, the location of the machinery and assembly lines makes it difficult to communicate using line-of-sight. Additionally, factory floors are typically rich radio signal scattering environments with a variety of tools and machines that affect shadowing and radio signal interaction in diverse ways. During production, both humans and robots move around carrying out their duties, creating a highly dynamic environment. Therefore, careful radio planning is required prior to installing a 5G system in a factory.
Depending on the industry's unique requirements, a private 5G network is the usual  
choice to meet the demands for high-performance, dependable, and secure networks for mission-critical applications. A public 5G network would be a preferable choice, if the enterprise merely needed a general-purpose network for basic communication applications.

The  integration of 5G wireless communication technology is anticipated to attain better outcomes in other vertical marketplaces such as healthcare, agriculture, supply chain logistics, and energy management. A key requirement for 5G industrial applications is access to the right radio spectrum, with the 3.5 GHz range being the most commonly used globally harmonized band. The wider selection of spectrum options available in 5G, particularly in licensed frequency bands, can meet a broader range of requirements and lower the cost of network equipment and devices. In modern manufacturing, where large amounts of data are generated, integration of various communication technologies is essential, with 5G technology providing not only high data rates and ultra-low latency to enable fast transmission and analysis of data, but the standard also specifies interfaces in the core network to integrate other wireless technologies such as WiFi and LoRa. An overview of the manufacturing use cases enabled by 5G is provided in Table~\ref{TABLE}.


\section{WiFi-7 for Industrial Communications}
\label{sec_wifi7} 
IEEE 802.11be, also referred to as WiFi-7, is the latest wireless local area network (WLAN) technology that can support the connectivity requirements of smart manufacturing in Industry 4.0 and beyond era. The technology offers several key features that make it a suitable choice for industrial environments, where real-time communication, security, and high-speed data transfer are critical.  Some of the key technological features of WiFi-7 are stated below, along with their applicability in smart manufacturing.

\begin{itemize}
    \item \textbf{High Modulation Order:} WiFi-7 utilizes Orthogonal Frequency Division Multiplexing (OFDM) with modulation order up to 4096-QAM, allowing it to transmit at very high data rates in a given bandwidth for vision based applications, for example. In the context of smart manufacturing, the significance of high data rates extends beyond the transmission of large volumes of data, as it also enables the ultra low latency transmission of small data packets within short timeframes, which is desirable in many industrial applications. Industrial processes that benefit from high data rate services include machine-to-machine communication, real-time monitoring, and production optimization, where the ability to swiftly transmit data is critical. It is worth noting that a signal-to-noise ratio (SNR) of approximately 40 dB is required at the receiver end to accurately decode a 4096-QAM signal, a threshold that may not always be attainable in many environments. However, antenna beamforming can help to alleviate this problem by increasing the channel gain. 
    

    \item \textbf{Multi-Link Operation:} WiFi 7 has an additional characteristic called Multi-Link Operation, which allows the access point and end devices to function concurrently over 2.4 GHz, 5 GHz, and 6 GHz frequency bands, thereby providing multiple channels for data transmission \cite{Deng_2020}, which aims to enhance network performance by increasing peak throughput, minimizing latency and jitter, and augmenting network reliability.  This makes sure that even if one connection fails, essential data will still be delivered, making WiFi 7 networks more dependable. On the other hand, link aggregation can also be performed to increase the network throughput significantly. In the industrial domain, these features are especially useful for processes such as machine-to-machine communication and real-time inventory control, where a dependable network is essential for proper and effective performance.

    \begin{figure}[t]
\centering
\includegraphics[width=\figwidth] {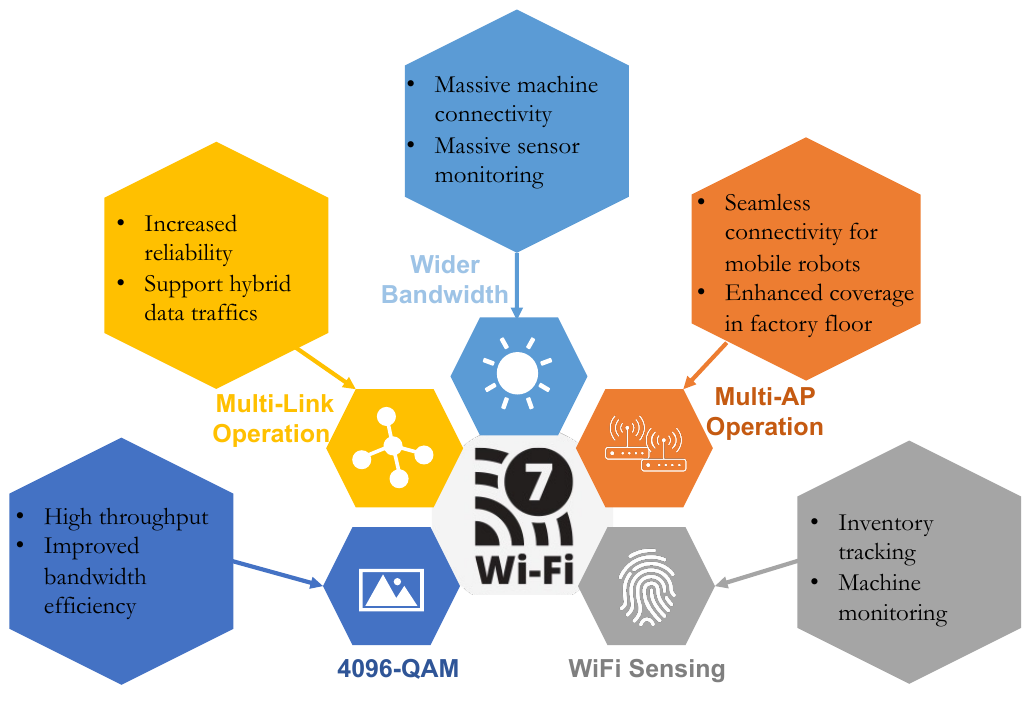}
\caption{Key features of WiFi 7 and their applications in smart manufacturing.}
\label{fig:wifi7}
\end{figure}

    \item \textbf{Wider Bandwidth:}  A distinguishing characteristic of WiFi 7 is its broadened bandwidth. After the initial adoption of 802.11ax, the WiFi industry is increasingly utilizing the 6 GHz band to swiftly enhance the peak throughput of WiFi, which will have a significant effect on industrial use cases. Consequently, conversations have arisen about the most optimal methods to make use of the available unlicensed spectrum, up to 1.2 GHz, located between 5.925 and 7.125 GHz, which more than doubles the bandwidth compared to what is available in the 5 GHz band alone \cite{8847238}.  By providing a wider bandwidth, WiFi 7 will have the potential to support a large number of industrial devices. Moreover, working in a less congested frequency spectrum also diminishes interference, which can be an issue in industrial settings where many systems and devices operate in close proximity.
    

\item \textbf{Multi-AP Operation:} WiFi 7's Multi-AP Operation permits multiple access points to come together and form a single, continuous network. This feature has the potential to facilitate seamless handover between WiFi networks and simplify the overall network configuration (e.g., selecting the channels for operation), and enhance the capacity of the WiFi network \cite(9152055).  By having multiple access points in sync, the coverage can be broadened across the entire factory floor, guaranteeing that all machines and mobile devices maintain a reliable and strong connection. Furthermore, cooperation among neighboring APs through the exchange of crucial scheduling information and channel state information (CSI) is a potential strategy to enhance the utilization of scarce radio resources \cite{Deng_2020}, particularly in an industrial environment with a high density of sensors and actuators, where co-channel interference can reach intolerable levels.

    

    \item \textbf{WiFi Sensing:} Wireless radio sensing is a cutting-edge feature of WiFi that allows WiFi networks to sense and detect the presence of people, objects, and other devices, even when they are not actively transmitting data. In smart manufacturing, WiFi sensing can have a range of important applications, including enabling location-based services, asset tracking, and improved safety and security. Location-based services will allow for real-time tracking of machines, devices, and personnel within the factory floor.  Asset tracking is another significant role for WiFi Sensing, making sure that costly machinery and equipment are not misplaced or stolen. By detecting the presence of these assets, manufacturers can monitor their usage, maintenance schedules, and movements, ensuring that they are always in good working condition and ready for use. Furthermore, WiFi sensing  can be a key to improving the safety and security of the smart factory \cite{Ma_2020}. By detecting people and machines, WiFi Sensing ensures a safe and secure work environment for all personnel by preventing accidents. WiFi Sensing, for instance, can alert workers of dangers in areas where machinery may pose a risk to personnel so that they can take appropriate measures.

\end{itemize}

The features of WiFi-7 and their usage in smart manufacturing is summarized in Fig.~\ref{fig:wifi7}. These features of WiFi-7 enable advanced smart manufacturing applications, including collaborative robots, augmented reality, and predictive maintenance. WiFi-7, however, is not without its limitations. Due to the use of higher modulation and higher carrier frequency,  the range of WiFi-7 signals will likely be reduced compared to previous generations operating at lower frequencies. This could be an issue in certain industrial environments where devices, machines, and systems are spread out over a large area.  In addition, since the standard has not yet been ratified, it is unclear when WiFi-7 access points and client devices will be available commercially.

\section{TSN for Industrial Communications}
\label{sec_tsn}
Real-time communication technologies that can provide guaranteed delivery of critical messages in a timely manner are of utmost importance in many manufacturing industries. Today, there exist a wide variety of wired and wireless communication technologies. However, industrial use cases that demand ultra-high reliability and low latency, such as control systems in automation applications, robotic motion control, safety-critical systems etc., rely on several dedicated and proprietary Ethernet protocols such as PROFINET, EtherCAT, Ethernet/IP, etc. \cite{TSN_survey_Seol}. These protocols, collectively known as Real Time Ethernet, are built upon Ethernet's legacy features, mainly designed to handle best-effort traffic. Time-sensitive networking (TSN) is a set of standards introduced to provide communication networks that demand real-time capabilities such as zero/extreme low packet loss and bounded end-end latency. In this section, we aim to provide a brief overview of the key elements of TSN that support the vision of smart manufacturing.

\begin{figure*}[t]
\centering
\includegraphics[scale = 0.5] {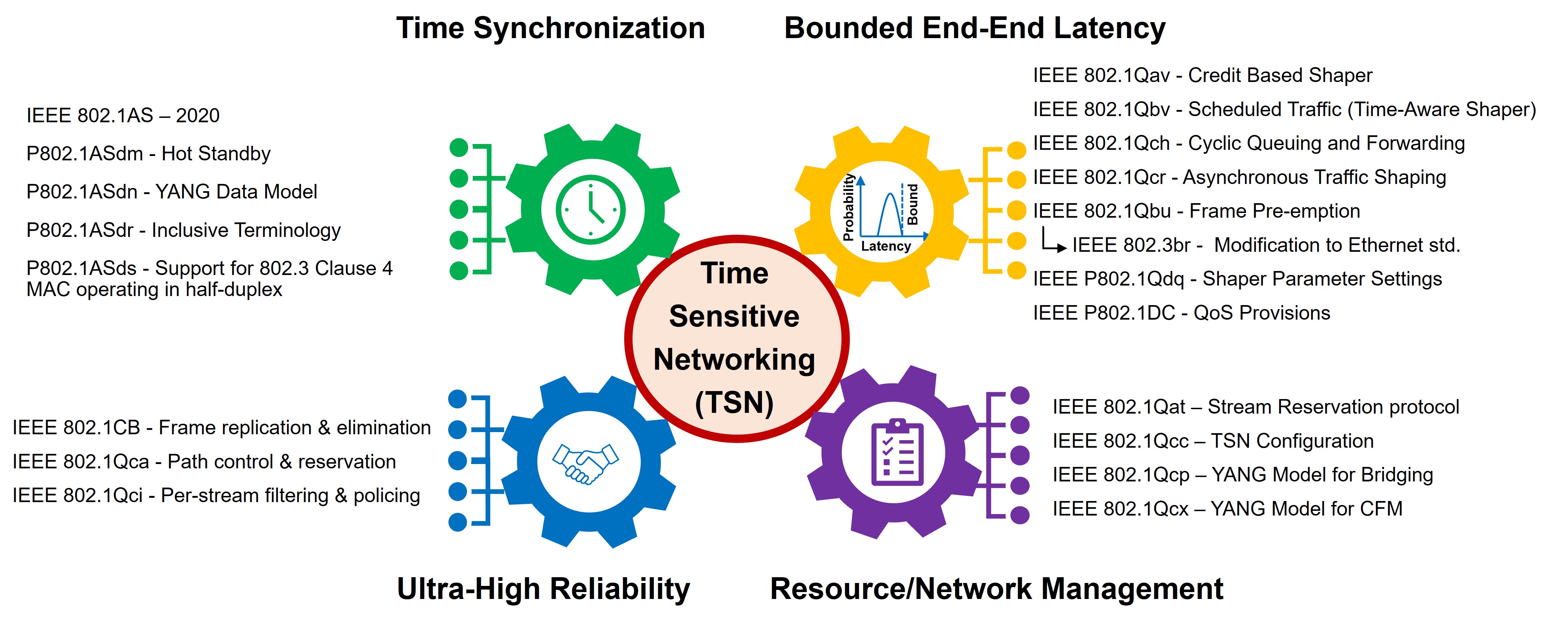}
\caption{TSN; Features, key mechanisms \& associated standards}
\label{fig_tsn}
\end{figure*}

TSN aims to achieve deterministic real-time communication through time-triggered message delivery in the network, where all the network elements are globally time synchronised. TSN tries to accommodate traffic flows that demand ultra-high reliability (zero-loss), low jitter and low latency on a bridged network, along with best-effort traffic. This is achieved by combining various mechanisms such as traffic shaping, defining various traffic queues and appropriate scheduling schemes, and reliability measures such as redundant/parallel message transmissions. Fig.~\ref{fig_tsn} outlines the main features/mechanisms associated with TSN, along with the respective standards. They can be broadly categorized into four pillars; 1) Time Synchronization, 2) Bounded Latency,  3) Ultra High Reliability, and 4) Network/ Resource Management \cite{perspective_IEEE, Finn_TSN}.

\textbf{Time Synchronization}: To enable strictly bounded end-end latency and several other features, TSN requires precise time synchronization up to sub-microsecond levels of accuracy among all the network elements such as switches, bridges, and end devices (manufacturing robots, PLCs etc.). TSN (wired) achieves this through the Precision Time Protocol (PTP – version~2) (IEEE 1588), using Ethernet frames in a distributed fashion to synchronize the timing of all the network elements to that of a master device (known as Grandmaster). The TSN time synchronization standard, IEEE 802.1AS (a subset of IEEE 1588), defines a generalized PTP (gPTP) profile to extend the time synchronization approaches beyond wired networks to wireless TSN. The currently active standard IEEE802.1AS-2020 \cite{ieee2020ieee} covers the protocols and procedures involved in selecting the timing source (best master) in the network, transport mechanisms for the synchronized time and indication of occurrence and magnitude of impairments (e.g., phase and frequency discontinuities). The time synchronization mechanism is based on a master-slave approach, where all the devices in the network are synchronized to the Grandmaster (the device with the most accurate time source in the network) through periodic exchange of sync and follow-up messages over a spanning tree. IEEE 802.1AS is an evolving standard, and details of active amendments (P802.1ASdm, P802.1ASdn, P802.1ASdr, P802.1ASds) can be found at \cite{ieee2020ieee}.

\textbf{Bounded End–to-end Latency}: In an industrial network, TSN adopts several mechanisms to provide the required end-to-end latency for different traffic classes (including time-critical messages). Special packet handling approaches such as traffic scheduling and traffic shaping techniques are introduced to guarantee bounded latencies. In the early days, the development of TSN standards considered Ethernet as the physical layer, and IEEE 802.1Q is a standard that supports VLANs on an Ethernet network. According to this standard, the IEEE802.1Q tag ("Priority Code Point" field – 3 bit) defines eight types of traffic classes having separate queues of their own per each Ethernet port. Various mechanisms (defined by IEEE 802.1Qav, IEEE 802.1Qbv, IEEE 802.1Qbu, IEEE 802.1Qch, and IEEE 802.1Qcr) \cite{Finn_TSN} can be used to achieve different QoS levels across the eight queues in multi-hop switched networks. IEEE 802.1Qbv (Time aware shaper) manages time-critical traffic flows based on a time-triggered scheduling approach. Each queue has a transmission gate that is opened/closed in a cyclic fashion for a particular time duration based on a gate control list (GCL). IEEE 802.1Qav (Credit based shaper) defines the operation of traffic flows with relaxed latency bounds. A ``credit" based transmission strategy is adopted to allocate the bandwidth (logically) to all traffic classes and thus prevents the starvation of lower priority traffic~\cite{atiq2021ieee, Seliem_IFIP}. IEEE 802.1Qbu (Frame Pre-emption) specifies the suspension/holding of an ongoing low-priority best-effort traffic (preemptable) transmission by a high-priority traffic flow (express traffic) to meet latency bounds. The associated IEEE 802.3br amendment defines two MAC interfaces: express and preemptable interfaces. Only a frame mapped to an express interface can pre-empt the frame mapped to a preemptable interface. IEEE 802.1Qch (Cyclic Queuing and Forwarding) discusses assigning message frames to egress queues according to their arrival time. IEEE 802.1Qcr (Asynchronous traffic shaper) also shapes the incoming traffic with the help of a ``credit" counter and schedules prioritized  over non-prioritized traffic, per hop, without the need for synchronization among various network elements. The amendment IEEE P802.1Qdq describes the recommended shaper parameter settings for bursty traffic needing bounded latency, and IEEE P802.1DC specifies the QoS features for a networked system rather than a bridge.

\textbf{Reliability}: This is another critical requirement for industrial networks in the smart manufacturing domain. Most communication technologies handle reliability through re-transmission mechanisms, which is not an option for time-critical traffic flows. IEEE 802.1CB (frame replication and elimination) ensures reliability in TSN networks through redundant frame transmissions across disjoint paths. IEEE802.1Qca (path control and reservation) handles the creation of multiple paths in the network by extending the application of Intermediate System to Intermediate System (IS-IS). It also specifies explicit path control, data flow redundancy, and distribution of control parameters for scheduling and time synchronization. IEEE 802.1Qci (per-stream filtering and policing) specifies the procedures for filtering frames of a particular data stream (using rule matching to filter streamIDs). It also addresses policing aspects to ensure that the systems/devices conform to the agreed configurations (e.g., allocated bandwidth) to guarantee QoS to different data streams.

\textbf{Resource/Network Management}: Proper configuration/management of various network elements (e.g., bridges, switches, end devices) and resources such as bandwidth, communication paths \& schedules, are crucial to achieving TSN's low latency, highly reliable communication capabilities. IEEE 802.1Qat (stream reservation protocol) deals with reserving resources and schedules between the source and the destination to achieve desired end-end QoS. IEEE802.1Qcc specifies three TSN network configuration models; fully centralized, hybrid and fully distributed \cite{TSN_review_mdpi}. Several data flows (streams) between end devices (acting as talker/listener) co-exist in a TSN. The ``User/Network Interface (UNI)" interface deals with the exchange of configuration information, such as join/leave request, resource configuration/release for a particular stream, or status response, between end users and the network. In a fully centralized model, end devices send their requirements to an entity called ``Centralized User Configuration (CUC)", which passes these details to the ``Centralized Network Configuration" (CNC) device. Accordingly, CNC manages all the streams and schedules in the network. In a hybrid model, the user's requirements are passed directly to the network, and CNC does the configuration without the existence of CUC in the network. In a fully distributed model, neither CNC nor CUC exits. The configuration data exchanges between bridges use YANG models (a data modelling language) as specified in IEEE 802.1Qcp. IEEE802.1Qcx specifies YANG models associated with fault management in the connectivity and currently both of these are part of the active IEEE 802.1Q-2022 standard\cite{ieee2022Q}.




\subsection{TSN over Wireless}

Although time sensitive networking is in principle independent of the MAC layer, all the above-detailed TSN mechanisms were specified over the last few years assuming Ethernet as the MAC layer. Their extension to the wireless domain is essential for smart manufacturing applications, and there is ongoing work to extend TSN mechanisms to operate over a wireless medium.
Emerging wireless technologies such as 5G and Wi-Fi 7 are expected to revolutionize the industrial sector, untethering several industrial applications, making it highly flexible, mobile and re-configurable at lower cost (installation and maintenance)\cite{ieee_stds_wifi, review_wireless}. Many of these applications demand deterministic latency and ultra-reliability, and hence TSN needs to be extended into the wireless domain since these applications/use-cases will be implemented over a hybrid network (having both wired and wireless devices), if not being completely wireless.

The wireless nature of the communication medium introduces several challenges. Dynamic link quality and packet error rates, especially in harsh industrial environments, shared medium access, interference due to other wireless transmissions, unlicensed spectrum, and mobility of the end-devices are some of them. In this section, we discuss how well the TSN capabilities can be extended / integrated with 5G and Wi-Fi-7, along with the ongoing efforts to meet the demands of the smart manufacturing industry using a hybrid wired-wireless TSN network.

\subsubsection{\textbf{TSN over WiFi}}
Wi-Fi is currently the most widely used wireless technology in the industrial domain, and the latest Wi-Fi generations, “IEEE 802.11ax / IEEE 802.11be”, have several features that enable the integration of TSN mechanisms. \textit{Precise time synchronization} among all the network elements is crucial in TSN. Time synchronization in 802.11 is achieved through the Fine Timing Measurement (FTM) protocol, an improvised version of the Timing Measurement (TM) protocol. In FTM, a station computes and adjusts its synchronization error and frequency drift with respect to its AP via repeated exchange of time-stamped FTM frames and their acknowledgements (ACK). Previous research shows that hardware-based timestamping can achieve precise time synchronization (10-40ns) in comparison with software-based time synchronization techniques ($1 \mu s$) \cite{atiq2021ieee}. According to IEC/IEEE 60802 (the TSN profile for industrial automation), the maximum tolerable synchronization error is $0.1-1 \mu$s for factory automation networks (from the grandmaster). HW-based time-stamping approaches using System-On-Chip solutions (e.g., FPGAs) are currently favoured, as SW timestamping on current COTS wireless cards/ modems cannot achieve the required time synchronization accuracy \cite{Val2022}. IEEE 802.1AS-2020 specifies the use of FTM for TSN integration over 802.11. It is essential to note that FTM message exchanges should not affect TSN traffic flows or vice versa.
In a Hybrid TSN network, wireless TSN can be at the edge of the network (acting as the last mile of the network) or act as a bridge between two wired TSN networks. Hence, it is essential to have smooth integration/translation of time synchronization across wired and wireless domains.

\textit{Traffic classification, shaping, and scheduling} mechanisms are vital in extending the TSN capabilities to the wireless domain. Industrial Wi-Fi networks must manage hard-real time, soft real-time and best-effort traffic through proper traffic classification methods. IEEE 802.11 (2016) supports traffic classification using VLAN tags employing traffic specification (TSPEC) and traffic classification elements (TCLAS), using the “traffic ID (TID)” field present in the IEEE 802.11 header. This is in accordance with the TSN traffic classifications specified in the IEEE 802.1Q standard and helps seamless integration between wired and wireless networks.

To support real-time traffic, efficient medium access control (MAC) is critical in 802.11 instead of the default contention-based MAC approaches. Several enhancements exist in IEEE 802.11ax and IEEE 802.11be to support time-critical traffic flows. The “Trigger Frame” (TF) based medium access scheme seems promising in comparison with other MAC mechanisms such as EDCA, HCCA, TWT etc., \cite{atiq2021ieee}. In addition, OFDMA and MU-MIMO capabilities help to reduce the contention and thus achieve low latency by enabling simultaneous transmissions to/from multiple users. Thus, the access point can deterministically schedule communications with the 802.11 devices in the network. As discussed earlier, a time-aware schedule (IEEE 802.1Qbv) defines the coordinated opening~/closing of gates so that the best-effort traffic does not interfere with the time-sensitive frames. Time-aware traffic scheduling (IEEE 802.1Qbv) can be achieved by reserving resources on top of efficient 802.11 MAC operation (e.g., TF-based access, EDCA). The multi-link operation capability in WiFi-7 (discussed in section \ref{sec_wifi7}) can help end devices to establish multiple links (thus multiple channel access opportunities) with the access point (AP). The newer 6GHz band will likely be preferred for time-critical traffic flows, while the congested 2.4/5GHz band can handle other traffic. Using the Stream Classification Service, an AP can allocate resources to the stations having time-critical flows within the BSS, and thus TSN traffic scheduling requirements (from 802.1Qbv) can be mapped onto 802.11be MAC. IEEE 802.11be also introduces “restricted target wake time (rTWT)” service periods (SP) scheduled by APs in a BSS for handling time-sensitive traffic, and every station is supposed to stop its transmission before rTWT SP. Multi-AP coordinated operation is another critical feature in 802.11be to allocate the channel resources efficiently in overlapping BSS to handle time-sensitive traffic.

Ultra-high reliability is another major challenge for wireless TSN networks due to the wireless nature of the medium. Reliability is mainly handled using redundancy mechanisms at intra-frame and inter-frame levels. In 802.11ax, the most reliable modulation and coding scheme (MCS) is MCS = 0 with BPSK (redundancy of $1/2$). Another scheme introduced in 802.11ax for better reliability is dual carrier modulation, where redundant data transmission occurs over different subcarriers. Inter-frame redundancy is achieved using ACK-based retransmissions or by sending extra copies of the same frame through different paths/resources without waiting for an ACK. Multi-link operation in WiFi-7 can enhance reliability through redundant transmissions over multiple links to a single AP or several APs. The multi-AP coordination feature helps to make use of redundant links across non-collocated APs. 

\subsubsection{TSN over 5G}

\begin{figure}[t]
\centering
\includegraphics [scale=0.5] {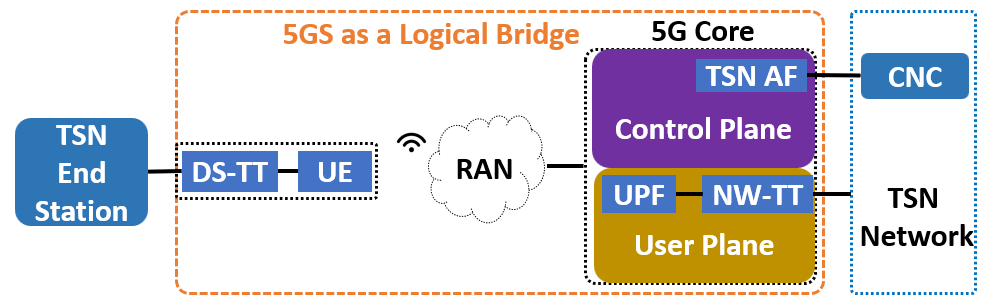}
\caption{5G system integration into TSN as a logical bridge}
\label{fig_5gtsn}
\end{figure}

 Private 5G networks and their capabilities, as discussed in section~\ref{sec_private5g}, are emerging wireless solutions for the smart manufacturing domain. Combining TSN capabilities with 5G helps to address the URLLC requirements of Industry 4.0. 3GPP Release 16 discusses the architecture (e.g., considers a fully centralized TSN configuration model) and functionalities required for 5G-TSN integration. The 5G system mainly consists of two components; a radio access network (comprised of UEs and gNBs) and a 5G core (5GC) network. The core network comprises different network functions associated with the user plane and control plane elements. 3GPP Rel 16 models the 5G system (5GS) as one or more logical/virtual TSN bridge(s) within the TSN network for 5G-TSN integration as shown in Fig~\ref{fig_5gtsn}. This logical 5GS TSN bridge has a TSN translator functionality to establish connectivity with the other elements in the network, both at the device side (DS-TT: Device side TSN Translator) and at the network side (NW-TT: Network side TSN Translator). DS-TT acts as an Ethernet port at the UE side user plane, whereas the NW-TT user plane, integrated with the 5GC user plane function (UPF), acts as an Ethernet port for the 5GS towards the external data network. NW-TT control plane is implemented as an application function (AF) and manages the interface towards CNC. The AF passes the TSN configuration/management-related information provided by the CNC to the 5GC (control plane). 	

5G supports TSN time synchronization defined by IEEE 802.1AS. 5GS itself is a time-aware system with an internal synchronization mechanism that keeps all the elements, such as gNB, NW-TT, DS-TT etc., synchronized to the 5G internal clock. TSN synchronization is another concurrent synchronization process in 5G-TSN integration. In this process, the TSN GM is outside the 5GS, and when a gPTP message (involved with the TSN synchronization discussed earlier) enters the 5GS through the NW-TT entity, it adds the ingress timestamp (in terms of 5GS time) and forwards it to the UE through UPF. When this gPTP message reaches DS-TT, with the help of an egress timestamp, the residence time in 5GS is calculated and included in the gPTP message and sent to the external TSN elements.  3GPP Rel. 17 talks about uplink TSN synchronization and UE-UE synchronization, where the TSN GM can be from the UE side. In this case, DS-TT acts as the ingress port and NW-TT as the egress port for the TSN gPTP messages.

5GS features several mechanisms for enhanced reliability when integrated with TSN. Some of them are multi-antenna transmissions, the use of multiple carriers, and packet duplication over different radio links in the 5G network. Redundancy in PDU sessions, UEs, and RAN dual connectivity at increased radio resource usage ensures reliability within the 5GS virtual bridge, an element of the TSN network. In addition to the user plane redundancy, the 5G core network supports redundancy by enabling the control plane network functions that can be deployed in multiple locations (e.g., edge, data centres).

Flexible numerology and configuration of OFDM signals are vital in enabling low latency in 5G networks. The shorter subframe slots and the mini-slots are the key enablers of low latency in 5G-TSN  systems. The 5G new radio also supports the pre-emption concept; URLLC transmissions can pre-empt non-URLLC data transmissions. Network slicing techniques and the 5G LAN-type service can be employed to handle time-critical TSN flows.

\section{A Comparative Analysis}

\begin{figure*}[t]
\centering
\includegraphics [scale=0.57] {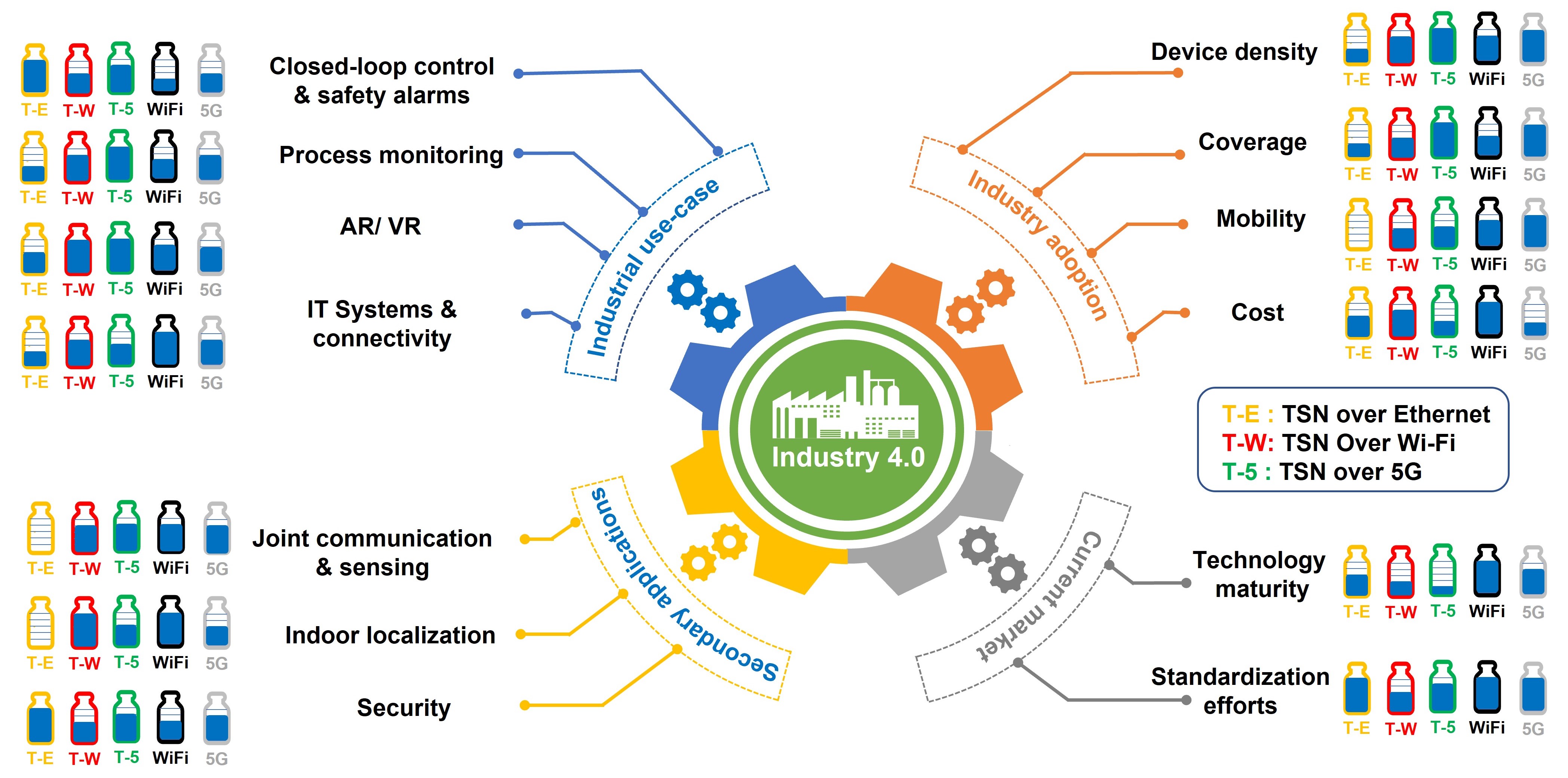}
\caption{Comparison of communication technologies for Industry 4.0 and beyond.}
\label{fig_comparison}
\end{figure*}

In sections \ref{sec_private5g}, \ref{sec_wifi7}, and \ref{sec_tsn}, we discussed the major features of 5G, WiFi and TSN networks (both wired \& wireless) that can enable smart manufacturing applications. In this section, we aim to provide a detailed comparative analysis of these technologies mainly from four perspectives; industrial use-cases, supported secondary applications, industry adoption and current market trends, as summarised in Fig.~\ref{fig_comparison}.

For the first dimension, we look at the industrial use-case perspective. There are a wide variety of use-case classifications in the literature \cite{5GACIA,review_wireless}, and here we capture a consolidated, broader application-oriented classification of them. Closed-loop control and safety-critical  alarms/ systems are the hard real time applications in smart manufacturing that demand ultra reliability and very low latency, and TSN over Ethernet (T-E) is the best solution. However, scenarios which need flexibility demand wireless communication options, and we know that the wireless medium always adds reliability/latency challenges. Among all the available wireless technologies, we believe TSN over 5G (T-5) has the advantage of combining the features of TSN over dedicated spectrum and hence it is better suited to handle such applications. WiFi is the least preferred choice due to the unlicensed spectrum implementation and associated interference and wireless channel contention issues.
Process monitoring use cases involve monitoring various parameters through different sensors. Such applications are relatively lenient in latency and reliability compared to closed-loop control applications, and the data-rate requirements generally tend to be small. However, wider coverage is essential as many processes in the process industry cover large industrial spaces, and wireless technologies would be a better choice than T-E as they eliminate wiring requirements. In these use cases, 5G provides a better range than WiFi, and the TSN integration provides additional reliability. Thus T-5 is better suited for process monitoring use-cases than T-W \& 5G, whereas WiFi seems the least preferred wireless technology for process monitoring.
MR/VR/AR are emerging technologies widely used in smart manufacturing to represent a digital 3D virtual factory for various applications, including training. These applications generally demand a very high data rate and flexibility to move around (hence T-E is least preferred). 5G  and WiFi both can provide higher data rates, and TSN integration offers additional capabilities for better reliability.

The second dimension focuses on the secondary services these technologies can provide in addition to their primary role of providing the communication infrastructure for manufacturing applications. Both 5G and WiFi (hence their TSN integrated versions) can offer sensing capabilities jointly with wireless communication. Indoor localization is another capability provided by both 5G and WiFi (hence their TSN-integrated versions). Localization using WiFi is relatively established in comparison with 5G-based localization. Both these features are not feasible over wired TSN. Security is another major factor that needs to be considered in connection with communication technologies. Wireless technologies are more prone to security threats compared to wired infrastructure. Among 5G \& WiFi, the unlicensed spectrum makes WiFi more exposed. 

The third dimension discusses the various factors smart manufacturing industries consider while adopting these technologies onto their production floor, in addition to the technical capabilities/offerings. 5G offers broader indoor coverage than WiFi (hence its TSN integrations), and TSN over Ethernet requires physical cabling to provision the T-E devices/switches. These aspects also play a role in device density. T-E requires an individual cable for each client. Although WiFi supports a possible bandwidth of 320MHz, a factory floor will have multiple APs with overlapping coverage areas. If an AP is operating with 320MHz bandwidth, the entire available spectrum is consumed and the nearby APs will use the same channels resulting in interference. Hence, it is not practical for each AP to make use of this wider channel. 5G mmWave supports multiple carriers of 100MHz, and carrier aggregation helps to obtain wider bandwidth, enabling dense device deployment. The smart factory floor consists of several mobile devices such as autonomous intelligent vehicles (AIVs), drones, etc.; another factor is how well these technologies support mobility. Although WiFi and 5G enable mobility, 5G provides smoother and more reliable handovers. The TSN integration creates challenges if devices are highly mobile, as the network configuration maintained by the CNC would need to be updated very often. Cost is another vital factor behind the industry adoption. WiFi is a more cost-effective solution than private 5G networks and hence their TSN integrated versions.

The fourth dimension that we consider is the latest trends in the market. Most enterprises adopt a technology once it is relatively mature and stable. WiFi is the most used wireless technology in smart manufacturing today, whereas private 5G networks are slowly seeing adoption. Several industries are currently carrying out private 5G proof of concept evaluations. The current lack of availability of native private 5G devices is a concern, even though a few Release 15/16 products are out in the market. Similar scenarios exist when it comes to TSN. Several products support TSN over Ethernet; however, TSN integration over wireless (5G \& WiFi) is in the early stages, which is also dependent on the standardization process. While the standards for T-E, WiFi and 5G are established, and newer releases are coming, the standardization process for TSN integration into WiFi and 5G is in the early stages. 

\section{Challenges and Future Directions}\label{sec:Chall}

\subsection{Dynamic network management}
Dynamic network management plays a pivotal role in enhancing communication performance within factory floors to deal with the dynamic and often unpredictable changes in the environment caused by the many metallic surfaces and the movement of various objects, including robots, automated guided vehicles, and other metallic objects as well as dense deployment of wireless devices. Dynamic network management is critical in facilitating optimal allocation of network resources, ensuring Quality of Service (QoS) guarantees, and adapting to changing network conditions, to enable efficient and reliable industrial operations.

The evolution of 5G technology is expected to focus on further enhancing network slicing capabilities to cater to the diverse requirements of industrial IoT use cases. This involves tailoring network slices to specific application needs, such as ultra-low latency, ultra-high dependability, and the ability to handle massive connections. Additionally, the integration of edge computing capabilities into 5G networks and the 5G LAN-type service offer the potential for low-latency processing and an improved user experience. In this context, dynamic network management and fast packet switching plays a critical role in efficiently allocating resources between the central cloud and edge nodes, thereby supporting latency-sensitive applications.

As WiFi 7 is anticipated to introduce improved multi-user MIMO methods, enabling multiple devices to transmit data simultaneously, dynamic network management becomes essential for optimizing scheduling and resource allocation for numerous users. This optimization process takes into account channel conditions, traffic patterns, and user priorities. To maximize spectrum utilization and alleviate congestion, existing spectrum access strategies, such as dynamic spectrum sharing or cognitive radio, may be adopted in WiFi 7 and new techniques need to be discovered. In this scenario, dynamic network management will facilitate the dynamic selection of the most suitable spectrum bands and their allocation to different WiFi services. Furthermore, future WiFi networks are expected to incorporate enhanced QoS methods to support a wide array of applications with varying requirements. Enhanced dynamic network management will be required to provide differentiated QoS based on application types, latency sensitivity, bandwidth demands, and traffic prioritization. The discovery of such adaptive approaches to network management will ensure the efficient utilization of resources and the seamless delivery of services tailored to the specific needs of different applications.



\subsection{Deployment issues}
The deployment of private 5G networks in manufacturing environments requires careful investigation, as traditional 5G deployment approaches may not suit the unique communication requirements of factory floors or processing plants. Hence, revisiting and rethinking base station placement policies is essential to tailor private 5G for diverse use cases in smart manufacturing.  On the other hand, upgrading to a new WiFi standard may necessitate the replacement or upgrade of old devices to support the new standard. Maintaining backward compatibility with older WiFi protocols can also be difficult.  As the number of WiFi devices grows, the available unlicensed spectrum may become congested, resulting in performance loss. Effective spectrum management and coexistence strategies will be critical for WiFi 7 implementation. Interference from other devices or neighbouring networks operating on the same or surrounding channels might degrade WiFi signals. For optimal implementation, proper channel allocation and interference mitigation strategies will be required. All of this requires novel research in deployment approaches for untraditional wireless environments and applications. 

TSN necessitates careful network device configuration and management to achieve accurate time synchronisation and prioritisation of time-sensitive traffic. This can be difficult, particularly with large-scale installations. TSN is based on standardised protocols and network device settings. During deployment, ensuring interoperability across different vendors' equipment, resource allocation and scheduling approaches can be difficult. TSN seeks to accommodate many forms of traffic on the same network infrastructure, including time-sensitive and standard Ethernet traffic. It can be difficult to achieve efficient network convergence while adhering to rigorous time requirements during deployment.  Integrating TSN into existing networks and legacy systems might be difficult since it may necessitate equipment modifications or replacements to support TSN's time-sensitive features. Research to novel discover methods for TSN network configuration and deployment are required.


\subsection{TSN-grade wireless performance}
To achieve TSN-grade performance over wireless communication channels, one must overcome several difficulties brought on by the special properties of wireless media and communication protocols. Due to interference and the stochastic nature of the channel, wireless systems typically have a much higher packet error rate (PER) than cable or optical fiber based communication systems.  TSN features like bandwidth scheduling and reservation are very good at delivering low latency/jitter over Ethernet. Such capabilities must take into account the wireless link characteristics such as achievable data rates and PER, which tend to change over time subject to radio environment changes and interference. The 3GPP Rel-15 defined Ultra-Reliable Low Latency Communications (URLLC) mode is a key 5G capability that enables TSN-grade performance. The URLLC mode, as explained in \cite{cavalcanti2019extending}, offers low latency  with high dependability for short packets in conjunction with the flexible 5G frame structure concept. Additionally, established QoS improvements for multiple concurrently active configurable grants and semi-persistent scheduling. Also, the extension of 802.1 TSN protocols over 802.11 is by default in line with the overall TSN reference architecture because 802.11 is one of the 802 LAN transport alternatives. However, in order for extensions of the 802.1 TSN protocols to function effectively, they must support the 802.11 MAC/PHY. 

Existing TSN services (such as scheduling implemented in network administration and configuration) and wireless networks (such as 802.11 and 5G) will require the definition of new abstractions and interfaces. To ensure predictable and dependable time-aware delivery of scheduled traffic across wired and wireless domains enabled by various wireless technologies, clear inputs, outputs, tasks, and responsibilities are required. The adoption and expansion of TSN networks over wireless networks will be facilitated by a standard interface to interact with wireless domains that abstract the underlying wireless connectivity technology. This work might be done in cooperation between the Avnu Alliance and IEEE 802.1 standards groups and could potentially include the definition of new wireless-specific parameters (for example, as an extension to the current 802.1Qbv YANG model).  This work could also influence wireless standard organizations, such as IEEE 802.11 and 3GPP.


%

\subsection{Implementation challenges}
New standards, such as 5G, WiFi 7, introduce new technologies. Theoretically speaking, there are several new features available with these new standards. But from a practical point of view, big efforts have to be made to implement them.  Careful testing and fine-tuning of these new technologies, especially in the factory floor environment, will be required. In the case of a factory floor environments, the implementation of new features can be particularly challenging. This is because factory floors are often complex and dynamic environments, and new features can have a significant impact on the way that work is done. As a result, it is important to carefully plan and test any new features before they are implemented in a factory floor environment. Also, purchasing and implementing new equipment, software etc. demands extra study and testing.  There is always a risk that new features will not work as expected, or that they will introduce new problems or security vulnerabilities.

\subsection{AI in industrial wireless communications}
We can design and implement practical communication technologies for various wireless channels in actual application environments when we have channel models that accurately depict the precise physical effects of these channels on transmitted radio signals. Traditional channel modellings techniques such as stored channel impulse responses, deterministic channel models and stochastic channel models need an in-depth understanding of the subject matter and technical proficiency in radio signal propagation modelling through electromagnetic fields. They are extremely complex with several parameters, unsuitable for predicting the statistical characteristics of wireless channels and not transferable to other communication settings.  To avoid these difficulties and complexities, machine learning and artificial intelligence approaches can help. For example, a generative adversarial network (GAN) framework is proposed in \cite{yang2019generative} to address the autonomous wireless channel modelling issue without requiring much theoretical investigation or data processing. Another example is the interplay between  RIS and AI in wireless communications, which has been explored in \cite{lin2020artificial}.

Modern manufacturing industries have progressed from electronic automation systems to industrial digital machine communication.
In order to gather and analyse data for guiding dynamic, intelligent systems in manufacturing, these systems need a high-speed connection. When 5G communications is combined with AI and big data, it becomes possible to analyse industrial data more quickly and make precise decisions based on the data. Smart factories provide better-connected machine ecosystems by using big data from interconnected devices powered by AI and 5G connectivity \cite{garg2019guest}. 

\section{Conclusion}\label{sec:Con}
In this paper, we have presented a comprehensive analysis of 5G, WiFi-7, and TSN in smart manufacturing applications for the Industry 4.0 and beyond era. The paper has highlighted the potential benefits and challenges of adopting these technologies in industrial networks, emphasizing the need for further research and development to address the issues of  reliability, compatibility and standardization. The comparative study of these technologies has provided valuable insights into their industrial use cases, secondary applications, adoption trends, and market opportunities. The paper's findings demonstrate that these technologies have significant potential to drive smart manufacturing towards the Industry 4.0 vision and beyond, and while each technology has advantages and disadvantages in regard to particular industrial applications, further efforts should be made to realize this potential. It is hoped that this paper will be a useful resource for researchers, engineers, and industrial practitioners interested in smart manufacturing and the role of emerging technologies in driving this transformation.

\bibliography{references}
\bibliographystyle{IEEEtran}

\end{document}